\documentclass{article}
\usepackage{amsmath,amsfonts,amscd,amsbsy}
\title{Maximum speed of quantum gate operation}
\date{}
\author{Lev B.\ Levitin, Tommaso Toffoli, Zachary Walton\\
{\footnotesize Boston University. ECE Dept., 8 Mary's St., Boston MA 02215}\\
{\footnotesize {\tt levitin@bu.edu}, {\tt tt@bu.edu}, {\tt walton@bu.edu}}}

 \makeatletter

 \DeclareRobustCommand\em
        {\@nomath\em \ifdim \fontdimen\@ne\font >\z@
                       \upshape \else \slshape \fi}

\def\@begintheorem#1#2{\sl \trivlist \item[\hskip \labelsep{\bf #1\ #2}]}
\def\@opargbegintheorem#1#2#3{\sl \trivlist
     \item[\hskip \labelsep{\bf #1\ #2\ (#3)}]}

 \mathchardef\BY="0202
 \def\@empty{}
 \newcommand{\asin}[2][]{{%extra braces for lower level, to forget \@cite
    \def\t@mp{#1}%
    \def\@cite##1##2{\marginpar{\hfil{\footnotesize$
    \ifx\t@mp\@empty\text{##2}\else\frac{\text{##1}}{\text{##2}}\fi$}\hfil}}%
\cite[#1]{#2}}}

 \newcommand{\sectlabel}[1]{\label{sect:#1}}
 
 \newcommand{\eqlabel}[1]{\label{eq:#1}}

 \newcommand{\Chapt}[2][]{\def\t@mp{#1}%
\chapter{#2} \ifx\t@mp\@empty\else\sectlabel{#1}\fi}
 \newcommand{\Sect}[2][]{\def\t@mp{#1}%
\section{#2} \ifx\t@mp\@empty\else\sectlabel{#1}\fi}
 \newcommand{\Subsect}[2][]{\def\t@mp{#1}%
\subsection{#2} \ifx\t@mp\@empty\else\sectlabel{#1}\fi}
 
 \newcommand{\Eq}[2][]{\def\t@mp{#1}%
\begin{equation}#2\ifx\t@mp\@empty\notag\else\eqlabel{#1}\fi\end{equation}}
 \newcommand{\Eqaligned}[2][]{\def\t@mp{#1}%
\begin{equation}\begin{aligned}#2\end{aligned}
\ifx\t@mp\@empty\notag\else\eqlabel{#1}\fi
\end{equation}}

 \newcommand{\sect}[1]{\S\ref{sect:#1}}      % Ref. to section or subsection
 \newcommand{\eq}[1]{(\ref{eq:#1})}	% Ref. to equation
 \long\def\endsubsection#1{\smallskip\hbox to\hsize{\leaders\hrule\hfill\ \sect{#1}}\medskip}

  \def\@arabic#1{\number #1} % my redefinition

\long\def\@makecaption#1#2{
	\vskip\abovecaptionskip
	\sbox\@tempboxa{{\small #1: #2}}%
	\ifdim\wd\@tempboxa>\hsize
	    {\small #1: #2\par}
	\else
	   \global\@minipagefalse
	   \hbox to\hsize{\hfil\box\@tempboxa\hfil}
	\fi
	\vskip\belowcaptionskip}

\def\figstrut#1{\hbox to\linewidth{\vrule height#1\hfill}}

\def\cstrip#1{\setbox0=\hbox{$#1$}\kern-.5\wd0\lower2pt\box0}
\def\rstrip#1{\setbox0=\hbox{$#1$}\kern-\wd0\lower2pt\box0}
\def\lstrip#1{\setbox0=\hbox{$#1$}\lower2pt\box0}
\def\tstrip#1{\setbox0=\hbox{$#1$}\kern-.5\wd0\lower\ht0\box0}
\def\bstrip#1{\setbox0=\hbox{$#1$}\kern-.5\wd0\raise\ht0\box0}
\def\Lstrip#1{\setbox0=\hbox{$\mskip2mu#1$}\lower2pt\box0}

\def\idpad{\thinspace}
\def\id{\idpad\begingroup \tt \let\do\@makeother \dospecials 
          \@ifstar{\@sid}{\@id}}
\def\@sid#1{\def\@tempa ##1#1{##1\endgroup\idpad}\@tempa}
\def\@id{\obeyspaces \frenchspacing \@sid}

\makeatother

 \newenvironment{tbmatrix}{\bigl[\begin{smallmatrix}}{\end{smallmatrix}\bigr]}
 \newcommand{\twomatrix}[4]{\begin{bmatrix}#1&#2\\#3&#4\end{bmatrix}}
 \newcommand{\ttwomatrix}[4]{\begin{tbmatciteiterix}#1&#2\\#3&#4\end{tbmatrix}}
 
 \newcommand{\ttwovector}[2]{\begin{tbmatrix}#1\\#2\end{tbmatrix}}

\def\H{\mathbf{H}}

\begin{document}

\maketitle

\begin{abstract}
\noindent We consider a quantum gate, driven by a general time-dependent
Hamiltonian, that complements the state of a qubit and then adds to it an
arbitrary phase shift. It is shown that the minimum operation time of the
gate is $\tau = \frac h{4E}(1+2\frac\theta\pi)$, where $h$ is Planck's
constant, $E$ is the average over time of the quantum-mechanical average
energy, and $\theta$ is the phase shift modulo $\pi$.
\end{abstract}

\bigskip\bigskip

\noindent It had been shown in \cite{margolus/levitin} that there exists a
fundamental limit to the speed of dynamical evolution of a quantum
system. Namely, the minimum time required for a system to go from a given
state to one orthogonal to it is
 \Eq[time-orth]{\tau=\frac h{4E},}
 where $h$ is Planck's constant and $E$ is the quantum-mechanical average
energy of the system. Expression \eq{time-orth} applies to the
\emph{autonomous} time evolution of a system, and it is not immediately
applicable to changes in the system state caused by the interaction with
another (external) system.

This paper considers the question of what is the minimum time of operation of
quantum gates that operate on qubits (i.e., quantum systems with
two-dimensional Hilbert space).

\medskip

Let
 \Eq[two_states]{\psi_1(0) = \ttwovector01\quad\text{and}\quad\psi_2(0) = \ttwovector10}
 be the two initial orthogonal stationary states of a qubit.  Consider a
``gate'' that complements the state of the qubit (a quantum inverter or a
controlled-{\sc not} gate with the controlling qubit in logical state 1) and
then adds to it an arbitrary phase shift $\theta$. This is a device that
applies an external interaction to the system for a certain time $\tau$ such
that at the end of this time interval
 \Eq[invert]{
	\psi_1(\tau)=\psi_2(0)e^{-i\theta}\quad\text{and}\quad
	\psi_2(\tau)=\psi_1(0)e^{-i\theta},
 }
 i.e., the two orthogonal states are swapped and a given phase shift $\theta$
is added to the resulting state.  Note that, owing to linearity, \eq{invert}
is a necessary and sufficient condition for an \emph{unknown} quantum state
$a\psi_1(0)+b\psi_2(0)$ of a qubit to be converted into the ortogonal state
with a phase shift $\theta$, provided that $\text{Re}(ab^*)=0$ (this
condition specifies a two-parameter family of states). Note also that the
overall phase of the state is essential, since this qubit is intended to be
part of a many-qubit system.

 \medskip

This problem was solved in \cite{levitin_delay} for the case of a
time-independent Hamiltonian. Here we treat the general case of a
\emph{time-dependent} Hamiltonian. \hbox{As in \cite{margolus/levitin}}, we
assume that the energy of the system is nonnegative (in other words, we
define energy relative to the energy of the ground state of the
system). Thus, the general form of the Hamiltonian we consider is
 \Eq[hami]{
	\H(t)= f(t)\twomatrix{E_{11}}{E_{12}e^{i\phi}}{E_{12}e^{-i\phi}}{E_{22}}
    	    = f(t) \H_0,
 }
 where $f(t)>0$ and $\H_0$ is a nonnegative definite self-adjoint
operator---which is equivalent to
 \Eq[implies]{
	E_{11}\geq0,\quad E_{22}\geq0,
	\quad\text{and}\quad E_{11}E_{22}-E_{12}^2\geq0.
 }

The time evolution of a system driven by a time-dependent Hamiltonian is
usually analyzed within the framework of perturbation theory (e.g.,
\cite{ll}). Here, however, we are interested in an \emph{exact} solution.

\medskip

Let
 \Eq[psit]{
	\psi(t) = a_1(t)\psi_1(0)+a_2(t)\psi_2(0);
 }
 then the Schr\"odinger equation for $\psi$ results in the following system
of differential equations for $a_1(t)$ and $a_2(t)$:
 \Eqaligned[schroed]{
  i\hbar\frac{da_1(t)}{dt} &= f(t)[E_{22}a_1(t)+E_{12}a_2(t)e^{-i\phi}]\\
  i\hbar\frac{da_2(t)}{dt} &= f(t)[E_{12}a_1(t)e^{i\phi}+E_{11}a_2(t)].
 }
 Let us introduce functions $b_1(t)$ and $b_2(t)$,
 \Eq[bees]{
	b_i(t)=k_ia_1(t)+a_2(t),\quad i=1,2,
 }
 where 
 \def\Radical{\sqrt{(E_{11}-E_{22})^2+4E_{12}^2}}

 \Eqaligned[kappas]{
	k_1&=\frac{e^{i\phi}}{2E_{12}}\left[E_{22}-E_{11}+\Radical\right]\\
	k_2&=\frac{e^{i\phi}}{2E_{12}}\left[E_{22}-E_{11}-\Radical\right].
 }

 The equations for the $b_i(t)$ are readily solved, yielding
 \Eqaligned[bsolved]{
	b_1(t) &= c_1 e^{-\frac{iE_1}\hbar F(t)},\\
	b_2(t) &= c_2 e^{-\frac{iE_2}\hbar F(t)}.
 }
 Here $F(t)=\int_0^t f(t)dt$; $c_1$ and $c_2$ are constants
depending on the initial conditions; and
 \Eqaligned[Econst]{
	E_1 &= \frac12\left[E_{11}+E_{12}+\Radical\right],\\
	E_2 &= \frac12\left[E_{11}+E_{12}-\Radical\right]
 }
 are eigenvalues of $\H_0$.

 From \eq{bees} we obtain
 \Eqaligned[aas]{
	a_1(t) &= \frac1{k_1-k_2}[b_1(t)-b_2(t)],\\
	a_2(t) &= \frac1{k_1-k_2}[k_1b_2(t)-k_2b_1(t)].
 }
 Now, let
 \Eq[psi1psi2]{
	\psi_i=a_{i1}(t)\psi_1(0)+a_{i2}(t)\psi_2(0),\quad i=1,2,
 }
 the initial conditions being
 \Eqaligned[aij]{
	a_{11}(0)=1,\quad & a_{12}(0)=0,\\
	a_{21}(0)=0,\quad & a_{22}(0)=1.
 }
 It follows from \eq{bsolved} and \eq{aij} that
 \Eqaligned[aijsolved]{
 a_{11}(t)&=\frac1{k_1-k_2}
	\left[k_1e^{-\frac{iE_1}\hbar F(t)}-k_2e^{-\frac{iE_2}\hbar F(t),}\right]\\
 a_{12}(t)&=\frac{k_1k_2}{k_1-k_2}
	\left[e^{-\frac{iE_2}\hbar F(t)}-e^{-\frac{iE_1}\hbar F(t)}\right],\\
 a_{21}(t)&=\frac1{k_1-k_2}
	\left[e^{-\frac{iE_1}\hbar F(t)}-e^{-\frac{iE_2}\hbar F(t)}\right],\\
 a_{22}(t)&=\frac1{k_1-k_2}
	\left[k_1e^{-\frac{iE_2}\hbar F(t)}-k_2e^{-\frac{iE_1}\hbar F(t)}\right].
 }
 Conditions \eq{invert} require that
 \Eq[atau]{
	a_{11}(\tau) = a_{22}(\tau)=0
 }
 and
 \Eq[ataubis]{
	a_{12}(\tau) = a_{21}(\tau)=e^{-i\theta}.
 }
 It follows from \eq{aijsolved} that, in order to satisfy \eq{atau},
one should have
 \Eqaligned[jumble]{
	&E_{11}=E_{22},\quad k_1=-k_2=e^{i\phi},\quad\text{and}\\
	&E_1=E_{11}+E_{12},\quad E_2=E_{11}-E_{12}.
 }
 Now, we obtain from \eq{atau} that
 \Eq[cosine]{
	\cos\left(\frac{E_{12}}\hbar F(\tau)\right)=0;
 }
 then \eq{ataubis} leads to the condition
 \Eq[halfcycle]{
  e^{i\phi}=e^{-i\phi},\quad\text{i.e.,}\quad\phi=n\pi,\quad n=0,1,2,\ldots.
 }
 Also, from \eq{ataubis},
 \Eqaligned[atauter]{
	a_{12}(\tau)=a_{21}(\tau)&=\pm ie^{-\frac i\hbar E_{11}F(\tau)}
				\sin\left(\frac{E_{12}}\hbar F(\tau)\right)\\
				&= e^{-i\theta},
} 
 where the plus sign corresponds to the choice $\phi=\pi$ and the minus
to the choice $\phi=0$.

From \eq{cosine},
 \Eq[sine]{
	\sin\left(\frac{E_{12}}\hbar F(\tau)\right)=\pm1.
 }
 Taking into account that $F(t)$ is a nonnegative and monotonically
increasing function of $t$, we conclude that, to achieve minimum $\tau$,
it must be that
 \Eq[pi2]{
	\frac{E_{12}}\hbar F(\tau)=\frac\pi2
 }
 and
 \Eq[either]{
 	\frac{E_{11}}\hbar F(\tau)=\begin{cases}
					\text{either}&\theta+\frac\pi2\\
					\text{or}&\theta-\frac\pi2
				\end{cases}.
 }
 Therefore,
 \Eq[eitherbis]{
	\text{either}\quad\frac{E_{11}}{E_{12}}=\frac{2\theta}\pi+1
	\quad\text{or}\quad\frac{E_{11}}{E_{12}}=\frac{2\theta}\pi-1,
 }
 the second choice being applicable only if $\theta\geq\pi$.

\medskip

Now, let us calculate the average over time, $E$, of the quantum-mechanical
average energy of the system:
 \Eq[energy_average]{
	E=\frac1\tau\int_0^\tau\langle\psi_i(t)|\H(t)|\psi_i(t)\rangle dt
	 =\frac{E_{11}}\tau F(t).
 }
 From \eq{pi2} and \eq{energy_average} we obtain
 \Eq[tau]{
	\tau=\frac{\pi\hbar}{2E}\cdot\frac{E_{11}}{E_{12}}
	=\frac h{4E}\cdot\frac{E_{11}}{E_{12}}.
 }
 Thus, to obtain the minimum value of $\tau$ we must take the smallest
ratio $E_{11}/E_{12}$. 

Finally, by \eq{eitherbis}, 
 \Eq[taufinal]{
	\tau=\tau(\theta)=\frac h{4E}\left[1+2\frac{\theta\bmod\pi}\pi\right].
 }
 It is remarkable that expression \eq{taufinal} is exactly the same as
that obtained in \cite{levitin_delay} for the case of the time-independent
Hamiltonian.

\bigskip

As a numerical example, consider experiments \cite{steane} made with Ca$^+$
ions in an ion trap. The characteristic wavelength of the transition between
the two relevant Ca$^+$ energy levels is $\lambda = 397\,\text{nm}$, which
yields $\tau = \tfrac \lambda{2c} \sim 6.62\cdot10^{-16}\,\text{s}$.

\bigskip

 Consider now a one-qubit quantum gate that makes an arbitrary unitary
transformation of a \emph{known} state such that the absolute value of the
inner product
 \Eq[minimum]{\big|\langle\psi(\tau_\alpha)|\psi(0)\rangle\big| = \cos\alpha.}
 A similar analysis shows that, for any $\alpha$ ($0\leq\alpha\leq\tfrac\pi2$),
the minimum time required for this operation is
 \Eq[final]{\tau_\alpha = \frac{\alpha h}{2\pi E} = \frac{2\alpha}\pi
\tau(0).}
 Expression \eq{minimum} is, again, exactly the same as in the case of the
time-independent Hamiltonian.

\medskip

It should be pointed out that the speed of quantum gates which have been
implemented up to now is very far from the fundamental limit
\eq{taufinal}. In particular, \cite{steaneetal} gives an excellent
theoretical analysis, as well as experimental results, of the operation time
of quantum gates that operate on trapped-ion qubits using laser pulses that
entangle the electronic and vibrational degrees of freedom of the trapped
$^{40}\text{Ca}^+$ ions. The gate time per ion obtained in \cite{steaneetal}
is of the order of $10^{-9}$\,s.

\end{document}